\documentclass[11pt,a4paper]{article}
\pdfoutput=1 
\usepackage{jinstpub} 
\usepackage{threeparttable}
\usepackage{subcaption}
\usepackage{tikz}
\usepackage{siunitx}
\graphicspath{{figures/}}
\usepackage{lineno}
\usepackage{hyperref}
\usepackage{lineno}
\usepackage{graphicx}
\usepackage[htt]{hyphenat}
\usepackage{makecell}
\usepackage{amsmath}
\usepackage{amsfonts}
\usepackage{float}
\usepackage{bm}
\usepackage{gensymb}
\usepackage{color}
\usepackage{todonotes}
\usepackage{comment}
\usepackage[symbol]{footmisc}

\usepackage[normalem]{ulem}

\title{\boldmath 
Performance of a SiPM-based, plastic scintillator muon veto prototype for CUPID}

\author[1]{M.~Moore,}
\author[1]{S.~Pagan,\footnote[2]{Contributions stem from research conducted as a graduate student and postdoctoral researcher at Yale University.}} 
\author[1]{E.~G.~Pottebaum,}
\author[1]{J.~A.~Torres,}
\author[1]{A.~Chizhik,}
\author[2]{R.~N.~Garcia,}
\author[1]{C.~Hidalgo,\footnote[4]{Now at Gran Sasso Science Institute, L’Aquila I-67100, Italy}}
\author[1]{K.~M.~Heeger,}
\author[1]{R.~H.~Maruyama,}
\author[1]{I.~Ponce,}
\author[1]{P.~T.~Surukuchi,\footnote[3]{Now at Department of Physics \& Astronomy, University of Pittsburgh, Pittsburgh, PA 15260, USA}}
\author[1]{J.~Wilhelmi}

\affiliation[1]{Wright Laboratory, Department of Physics, Yale University, New Haven, CT 06520, USA}
\affiliation[2]{Physics Department, California Polytechnic State University, San Luis Obispo, CA 93407, USA}

\emailAdd{jorge.torresespinosa@yale.edu}
\emailAdd{emily.pottebaum@yale.edu}

\date{\today}

\abstract{CUPID will be a next-generation experiment searching for neutrinoless double beta decay in the inverted mass ordering regime. The reduction of backgrounds in the region of interest is critical to the performance of the experiment. Despite its underground location, muon-induced events will be a non-negligible source of background for CUPID, and their mitigation will be critical in reaching CUPID's target sensitivity. This mitigation will be achieved with a muon veto system, which must fit within the physical constraints of the existing infrastructure while maximizing geometrical coverage. We present the design, construction, and characterization of prototypes for a modular system of plastic scintillator panels with embedded plastic wavelength-shifting fibers connected to silicon photomultipliers for the CUPID muon veto. The 100$\, \times \,$50$\, \times \, $2.5\,cm$^3$ panel prototype presented here exhibited a light yield of ($55.9 \,\pm \,1.5$)\,p.e./MeV with a position reconstruction resolution of 25$\,\times \,$25\,cm$^2$. This design also achieves a muon detection efficiency of $(98\,\pm \,1)\%$. We compare the light yield, uniformity, and position reconstruction potential of different prototype designs.}


\begin{document}

\maketitle
\flushbottom
\raggedbottom

\section{Muon-induced backgrounds in CUORE and CUPID}

Neutrinoless double beta decay ($0\nu\beta\beta$) is a theorized ultra-rare decay process that, if observed, would confirm the Majorana nature of neutrinos~\cite{PhysRevD.25.2951}. Experiments searching for $0\nu\beta\beta$, such as the Cryogenic Underground Observatory for Rare Events (CUORE)~\cite{NatureCUORE2022, CUORE:2017tlq}, are currently operating or completed, and the next generation of tonne-scale experiments is in development. CUORE's successor will be the CUORE Upgrade with Particle Identification (CUPID)~\cite{CUPID:2025avs} experiment. CUORE is an array of 988 TeO$_2$ crystal cryogenic calorimeters located at the Laboratori Nazionali del Gran Sasso (LNGS)~\cite{NatureCUORE2022}, planned to continue data taking until 2026. Following CUORE, CUPID will be commissioned in the same experimental infrastructure with upgrades. CUPID will be an array of  Li$_2$$^{100}$MoO$_4$ scintillating crystal cryogenic calorimeters. LNGS is a large underground research facility under the Gran Sasso mountain in central Italy, which provides 3600 meter-water equivalent (m.w.e). of rock shielding. This overburden heavily suppresses the muon flux--by about 6 orders of magnitude--from approximately $1\, \times \, 10^{-2}\,\text{cm}^{-2}\,\text{s}^{-1}$ at Earth's surface to $2.6\, \times \, 10^{-8}\,\text{cm}^{-2}\,\text{s}^{-1}$ underground at LNGS~\cite{Cecchini_2009,Bellini:2011yd,Mei:2005gm}. This enables the operation of ultra-low background experiments searching for rare events like $0\nu\beta\beta$. 


 
 CUPID will fully explore the inverted mass hierarchy region and part of the normal hierarchy region of the effective Majorana mass parameter space. To achieve this, CUPID will employ new technologies and a different isotope ($^{100}$Mo) with a higher $Q$-value ($Q_{\beta\beta} = 3034.40(17)$\,keV)~\cite{Rahaman:2007ng} to decrease the background index (B.I.) in the region of interest (ROI) by about two orders of magnitude compared to CUORE, i.e.,~from ${\sim}10^{-2}$\,counts/(keV\,kg\,y) to ${\sim}10^{-4}$\,counts/(keV\,kg\,y)~\cite{NatureCUORE2022,CUPID:2025avs}. To do this, CUPID will employ a dual-readout technology with heat and light channels to achieve $>99.9\%$ discrimination of alpha events~\cite{Poda:2021}, which constitute $\sim90\%$ of the backgrounds in CUPID's ROI. After alpha rejection, the primary leftover sources of background radiation are radioactive contamination in the detector and its surrounding structures, as well as pile-up events from the two-neutrino double beta decay (2$\nu\beta\beta$) of $^{100}$Mo. However, cosmogenic muons also represent a non-negligible source of backgrounds, even with the attenuation from the LNGS rock overburden. Muons produce prompt background events by directly depositing energy in the detector, or by producing secondary particles from muon-induced spallation, yielding a rate of approximately 2\,events/hour or ${\sim}0.6$\,mHz in CUORE~\cite{CUORE:2024fak}. Muon-induced events yield a background index of ${\sim}10^{-4}$\,counts/(keV\,kg\,y) in the ROI for $0\nu\beta\beta$ in $^{100}$Mo~\cite{CUPID:2025avs}. This means that without any sort of mitigation, muon-induced events alone would saturate all of CUPID's background budget. An active muon veto will be included in CUPID to reduce this background, with the requirement that muons in the ROI be rejected with an efficiency of at least $98\%$. 

In this work, we focus on the design and performance of prototype panels for the muon veto system. The design of the CUPID muon veto system is driven by the need to optimize the muon-tagging efficiency while fitting within the existing spatial constraints of the CUORE/CUPID infrastructure. The background at LNGS for this system is dominated by environmental gamma rays~\cite{GammaBackgroundLNGS}, particularly those at $\sim$2.6\,MeV from the decay of $^{208}$Tl, the gamma ray closest in energy to the muon signal region of $\sim$5\,MeV for our design. The prototype panels make use of photodetectors that are themselves a source of background due to dark noise, which arises from random fluctuations of thermally-generated electrons within individual pixels of each photodetector.

\begin{figure}[ht!]
  \centering
  \begin{subfigure}[b]{0.4\textwidth}
    \includegraphics[scale=0.87]{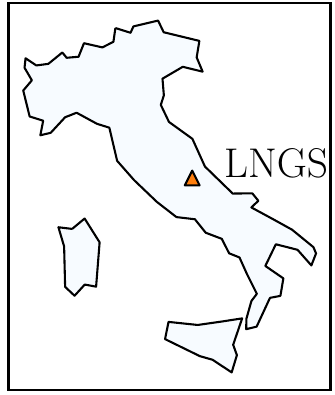}
  \end{subfigure}
  \begin{subfigure}[b]{0.45\textwidth}
    \includegraphics[scale=0.5]{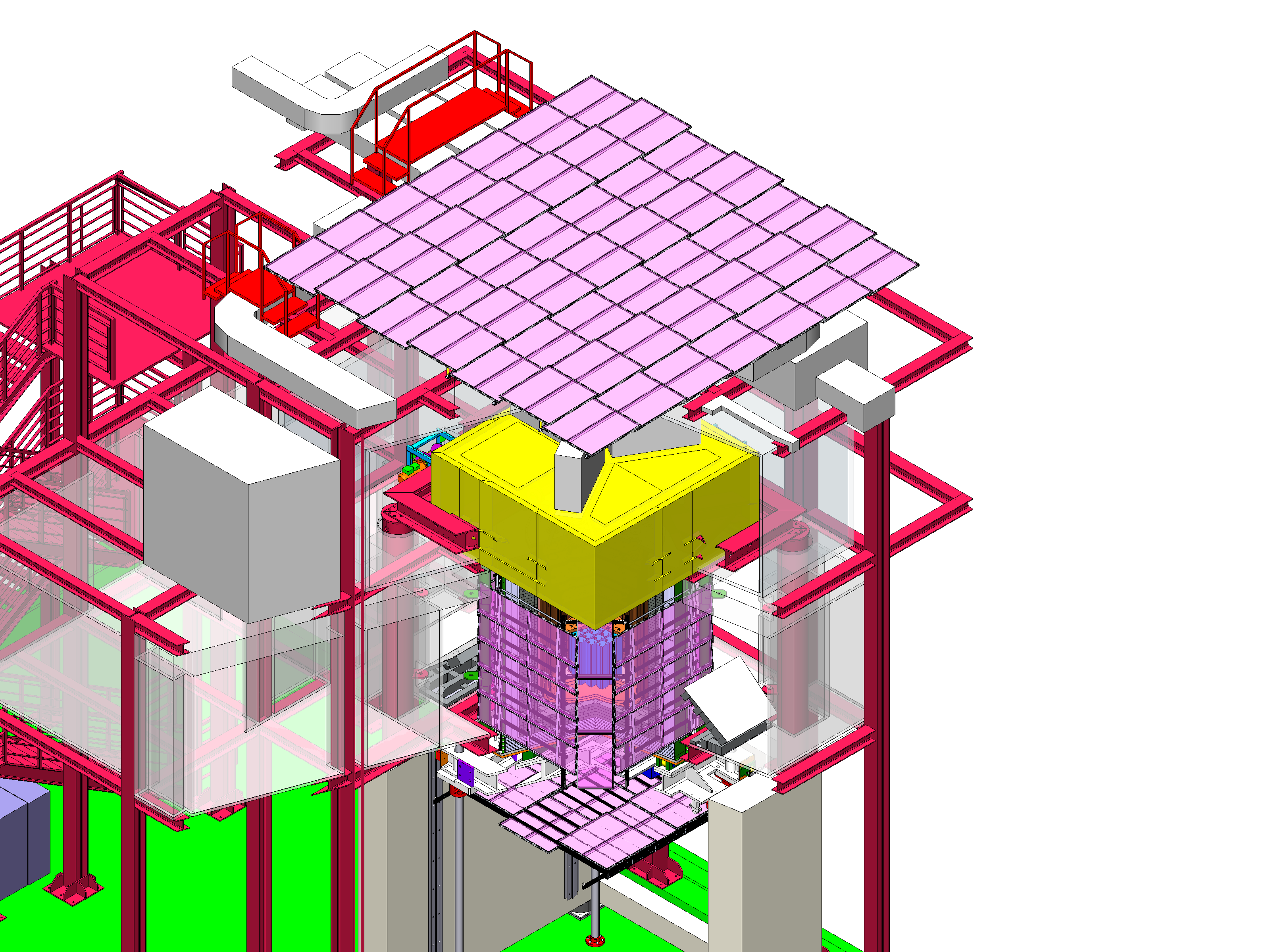}
  \end{subfigure}
  \caption{\label{Stencil} Left: Location of LNGS in Italy. Right: Diagram of CUORE detector and infrastructure with possible placements of the muon veto system shown in pink above, below, and around the sides of the detector. After upgrades, CUPID will be housed in the same infrastructure.}
\end{figure}

CUORE was built and has been operating without an active muon veto system. To achieve the two orders of magnitude reduction in background required for CUPID, a muon veto system must be added within CUORE's infrastructure. There are a few potential placements for the panels in the given infrastructure, one of which is shown in figure~\ref{Stencil}. Performing track reconstruction with the muon veto system is also desirable, but not required. This capability would complement muon track studies and other physics analyses that rely on events in multiple crystals. Based on these requirements, the key goals of this study were improving light yield, optimizing the design for underground environments, and assessing detection efficiency with different panel configurations.


The proposed muon veto system consists of single modules of wavelength-shifting (WLS) fibers embedded into a plastic scintillator panel and connected to silicon photomultipliers (SiPMs) as light detectors. The combination of plastic scintillator, WLS fiber, and light detectors is common for muon veto devices~\cite{OMineev_2011, BUGG201491, ZONG201882, Luo:2023inu, AHARONIAN2021165193}. This work describes the development, characterization, and testing of prototype modules for the CUPID muon veto.

\section{Prototype design and fabrication}

The prototypes described in this paper were developed to study the light yield, background discrimination, and muon tracking capabilities of different panel designs. Multiple aspects of the panels' design and fabrication were tested, such as fiber design, machining techniques, wrapping methods, and coupling to light detectors. To increase the light yield and uniformity, panels were embedded with WLS fibers in a swirly pattern that utilizes a high fiber density and minimizes tight turns where light can be lost~\cite{volchenko2008level}. In systems using similar materials, increasing the WLS fiber density has been shown to improve light yield~\cite{SEO2022167123Amore, BUGG201491, Lu:2023sbi}. SiPMs were chosen as light detectors to be coupled to the ends of the WLS fibers due to space constraints in the CUORE infrastructure. In larger prototypes, multiple spiral patterns were included with the motivation of creating a finer spatial resolution for tracking muons within modules. A single panel module operated with a self-trigger and a double panel module operated with a coincidence trigger were also fabricated to test how this could aid in discriminating between muons and gamma backgrounds. Multiple other variables were considered to meet the requirements of these panels and are detailed in this section. 

\begin{figure}
\centerline{\includegraphics[scale=.23] {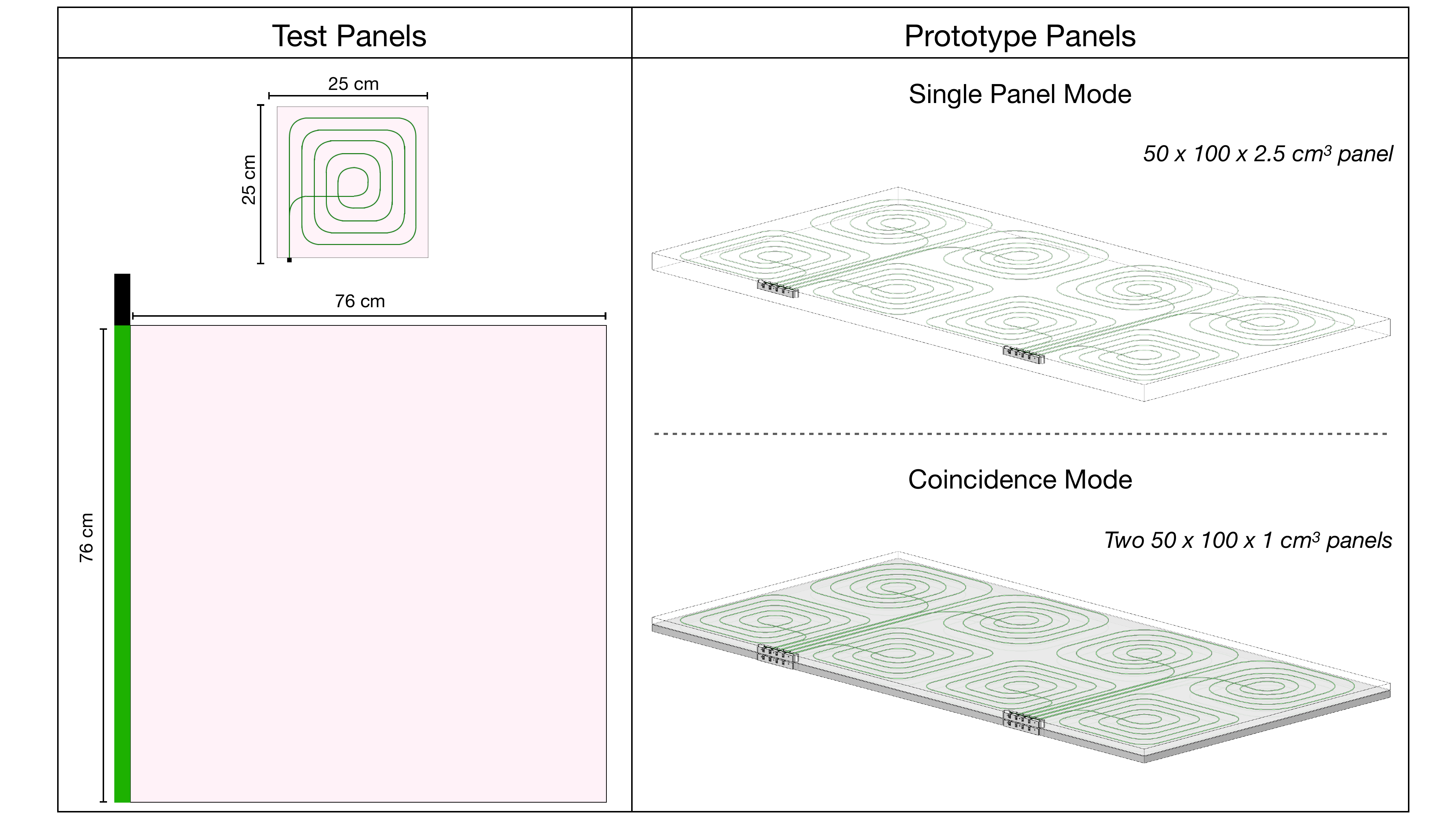}}
\caption{\label{prototypePanels} Muon veto panel prototype designs. The fiber design for the Small Panel, shown in top view, is read out by a single SiPM (top left). The top view of the LBNL panel shows the WLS plastic strip (green) and PMT (black) on the left side of the panel (bottom left). Fiber designs for Prototypes 1 (top right) and 2 (bottom right), consisting of eight "mini-modules" read out respectively by 8 SiPMs at the front of the panel, are shown from a top-angled side view. Prototype 1 is a self-triggered single panel module which utilizes an \texttt{OR} trigger for each SiPM readout channel. Prototype 2 is a coincidence-triggered double panel module, in which two panels are stacked and SiPM readout channels for overlapping mini-modules are sent to an \texttt{AND} trigger.}
\label{MV_prototypes}
\end{figure}

During initial prototyping performed at Yale's Wright Laboratory, fabrication methods were tested on a 25$\, \times \,$25$\, \times \,$2.5\,cm$^3$ panel (the "Small Panel" prototype). From the testing and characterization of this prototype, two different sizes of larger single module prototypes were developed. The prototypes include a single panel module ("Prototype 1") and a double panel module ("Prototype 2") using panels of size 100$\, \times \,$50$\, \times \,$2.5\,cm$^3$ and 100$\, \times \,$50$\, \times \,$1\,cm$^3$, respectively (see figure~\ref{MV_prototypes}). A panel that includes a wavelength-shifting bar and a photomultiplier tube (PMT) was also characterized. This panel was fabricated to be used in a muon veto system at the Low Background Facility at the Lawrence Berkeley National Laboratory (LBNL) and was also studied to compare uniformity and light yield~\cite{LBNLMuonVetoPanels}. This panel is about 10 years old, and therefore its performance could also be a proxy for the aging properties of the scintillator material in our design, since they both were procured from the same producer, Eljen Technologies~\cite{eljenEJ200}.

\subsection{Material selection and fabrication process}

All panel prototypes were made with EJ-200 plastic scintillator from Eljen Technologies~\cite{eljenEJ200}. 1\,mm diameter, round BCF-92 WLS fibers from St.~Gobain were embedded in plastic scintillators for Prototypes 1 and 2~\cite{Luxium}. The Small Panel used 1\,mm diameter Y-11(200)M WLS fibers from Kuraray~\cite{kuraray}. To increase the efficiency of light collection, the panels were wrapped in DuPont Tyvek 8740D secured with Tyvek Tape~\cite{Tyvek,TyvekTape}. From tests wrapping small pieces of plastic scintillator, it was found that two layers of DuPont Tyvek 8740D resulted in improvement of light yield over one layer. To block out ambient light, the panels were also wrapped in one layer of aluminum foil and then in a layer of high-performance black masking tape~\cite{BlackTape}. These layers serve as reflectors and light blockers to make the panels light-tight.



Before scaling to the full 100$\, \times \,$50\,cm$^2$ surface area,  25$\, \times \, $25\,cm$^2$ panel prototypes were machined and tested at Yale's Wright Laboratory. To machine grooves for WLS fiber, the plastic scintillator was submerged in a heat bath of soapy water~\cite{eljenMachiningPolishing}. A programmable Computer Numerical Control (CNC) machine was used to cut smooth, 4\,mm deep grooves. To embed the WLS fiber, UV-curable Norland Optical Adhesive 68 was applied where the WLS fiber exited the panel and in small dots along the path of the fiber~\cite{NOA68}. Lastly, EJ-500 optical cement was used to permanently attach and optically couple the fiber to the plastic scintillator~\cite{eljenEJ500}. During this process, it was found that consistent fiber positioning was difficult to achieve and also critical for the alignment of the SiPM with the fibers. A fiber stencil and counterweight were designed to achieve uniform WLS fiber placement in Prototypes 1 and 2. The stencil and weight secured the fiber at the bottom of the panel grooves so that they could be aligned with the SiPM.





Eljen Technologies in Sweetwater, TX fabricated the full-size scintillator panels and embedded WLS fibers in the panels. Eljen used the fiber stencil and counterweight, along with optical adhesive NOA 61 and EJ-500 to secure the fibers.  Small cutouts were milled on the scintillator panel at the ends of the fiber to create space for custom PCB SiPM mounts. 

\subsection{Prototype designs} 
\label{sec:Prototypes}

In this section, we describe the designs of the four prototypes of interest to this paper: the Small Panel, Prototype 1, Prototype 2, and the LBNL Panel. A summary of each design is provided in  table~\ref{Table:Prototypes}.

\begin{figure}[H]
  \centering
  \begin{subfigure}[b]{0.4\textwidth}
    \includegraphics[scale=0.47]{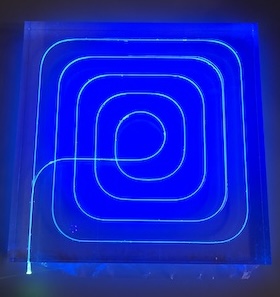}
  \end{subfigure}
  \begin{subfigure}[b]{0.45\textwidth}
    \includegraphics[scale=0.08]{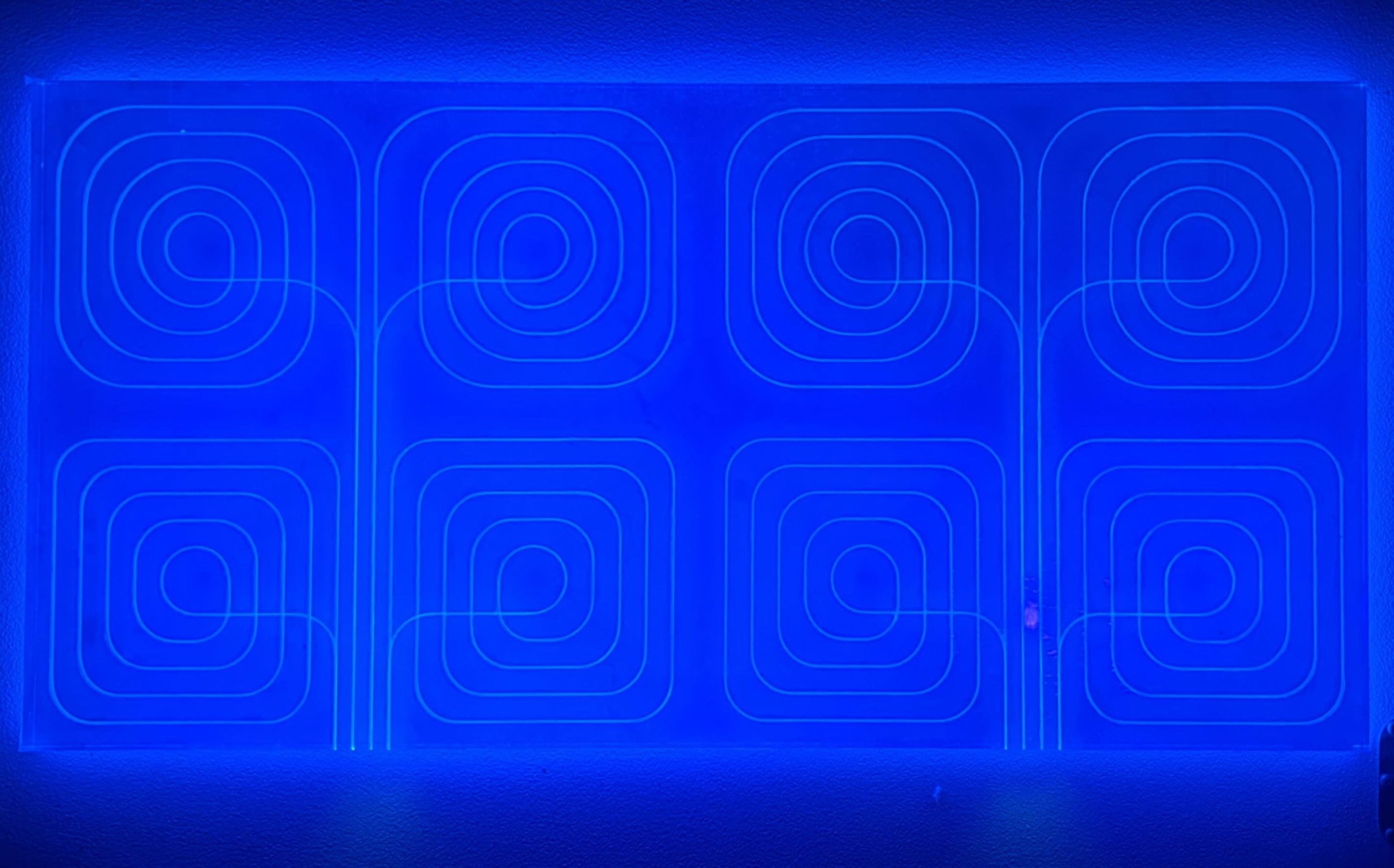}
  \end{subfigure}
  \caption{\label{SmallPanel} The Small Panel (left) and Prototype 1 panel (right) under UV light. The wavelength-shifting fiber pattern of the mini-modules is clearly visible with this view.}
\end{figure}

\paragraph{Small Panel}
The first prototype was made from a single, self-triggered panel of plastic scintillator. The WLS fiber design included 16 quarter turns of a 3\,cm bending radius and 4 quarter turns of a 2.5\,cm bending radius. The grooves of the design were 4\,mm deep, and the straight segments of fiber in the spiral are 2\,cm apart. This bending radius is smaller than recommended by the manufacturer, however the light loss due to all bends in this pattern is estimated to be below 20\% from the specifications \cite{kuraray}. The chosen geometry was a compromise that ensured good performance in terms of light yield and uniformity. This layout is shown in figure~\ref{prototypePanels} and is shown under UV light in figure~\ref{SmallPanel}. 

\paragraph{Prototypes 1 and 2}
Two different WLS fiber patterns were tested in Prototypes 1 and 2. In Prototype 1 the pattern consists of 16-quarter turns of a 3\,cm bending radius and 4-quarter turns with a 2.5\,cm bending radius in a 25$\, \times \,$25\,cm$^2$ panel area. This design is shown in the top right of figure~\ref{prototypePanels} and is shown under UV light in figure~\ref{SmallPanel}. Prototype 2 uses a different fiber pattern which consists of 12-quarter turns of a 5.5\,cm bending radius, 4 turns with a 4.5\,cm bending radius, and 4 turns with a 2.5\,cm radius in a 25$\, \times \,$25\,cm$^2$ panel area. This design is shown in the bottom right of figure~\ref{prototypePanels}. The 8 swirls, or "mini-modules," in each prototype read out to a distinct photodetector. We use 13360-3050PE Hamamatsu SiPMs~\cite{MPPCS13360-3050PE} for light detection, which have a collection area of 3$\, \times \,$3\,mm$^2$. Two fiber ends feed into each SiPM. The bending radius follows the manufacturer's guidance for the minimum bending radius \cite{StGobain_private}. The goal of this pattern was to enable a spatial resolution of approximately 25$\, \times \,$25\,cm$^2$ on tagged muons while maintaining a large module size. Larger modules are desirable for installation and operation as long as a high light yield can be maintained.    

Prototypes 1 and 2 also varied in thickness. The 2.5\,cm Prototype 1 module is designed to operate as a single, self-triggered panel, and 1\,cm thick Prototype 2 panels are stacked together and operated as a double panel module. An event in both the top and bottom panel is needed to trigger Prototype 2, which was designed to reduce gamma backgrounds and detect through-going muons with a thinner module. Both self-triggered and coincidence-triggered modules of the same fiber layout were fabricated to test background and muon discrimination in both modes.

\paragraph{LBNL Panel}
In addition to the new prototypes described above, this paper includes results from a muon veto panel used in the Low Background Facility at LBNL. The 76$\, \times \,$76$\, \times \,$2.5\,cm$^3$ LBNL Panel was designed to be simple, stable, and easy to operate. It was made from EJ-200 plastic scintillator, which was diamond milled on all sides except for a frosted edge that diffuses light. A thin strip of EJ-280 wavelength-shifting plastic was attached to the frosted edge side of the panel. A Hamamatsu R1924A photomultiplier tube was coupled to the end of this WLS plastic bar with an EJ-560 silicon rubber optical interface pad~\cite{eljenEJ560}. A schematic of this design is shown in the bottom left of figure~\ref{prototypePanels}. The panel was wrapped in reflecting foil and black vinyl light-tight outer wrap. This panel was manufactured in 2012, and it has been located at the LBNL Low Background Counting Facility since fabrication. These panels were shipped from LBNL to Wright Lab and re-characterized in early 2023. 

\captionsetup[subfigure]{labelformat=empty}

\begin{center}
\captionsetup{justification=raggedright, singlelinecheck=false}
\begin{table}[H]
\caption{Summary of the configuration and properties of the four panel prototypes. Note that Prototypes 1 and 2, and the Small Panel, use a SiPM readout whereas the LBNL Panel uses a PMT readout.}
\label{Table:Prototypes}
\begin{tabular}{ | p{1.95cm}| p{2.3cm} | p{1.9cm} | p{2.1cm} | p{1.7cm} | p{1.9cm} |} 
 \hline
\textbf{Prototype} & \textbf{Size (cm$^3$)} & \textbf{Trigger Mode} & \textbf{WLS Fiber} & \textbf{Fiber Bending Radii (cm)} & \textbf{Coupling} \\

 \hline
1 & 100$\, \times \,$50$\, \times \,$2.5 & \makecell[l]{self \\ trigger} & BCF-92 & 2.5, 3.0 & mechanical \\ 

\hline
2 & \makecell[l]{2 panels, \\100$\, \times \,$50$\, \times \,$1} & \makecell[l]{coincidence \\trigger} & BCF-92 & \makecell[l]{2.5, 4.5, \\ 5.5} & mechanical  \\ 

 \hline
\hline

\bfseries\makecell[l]{Test \\ Panels} &  &  &  & &  \\
\hline
Small Panel & 25$\, \times \,$25$\, \times \,$2.5 & \makecell[l]{self \\ trigger} & Y-11(200)M & 2.5, 3.0 & air  \\ 
\hline
LBNL & 76$\, \times \,$76$\, \times \,$2.5  & \makecell[l]{self \\ trigger} & \makecell[l]{WLS \\ Plastic Strip \\ (EJ-280)} & N/A & \makecell[l]{silicon \\rubber \\optical \\interface \\(EJ-560)}  \\ 
 \hline
\end{tabular}
\end{table}
\end{center}

\section{Experimental setup}
\label{sec:ExperimentalSetup}

\subsection{SiPM coupling and quality assurance}


In Prototypes 1 and 2, and the Small Panel, SiPMs were used as light detectors to create a compact, scalable system. These prototypes used 13360-3050 Hamamatsu SiPMs~\cite{MPPCS13360-3050PE, MPPCS13360-3050CS}. Low dark noise and crosstalk rates were the primary considerations when choosing the SiPM model. The crosstalk probability of 13360-3050 Hamamatsu SiPMs is ${\sim}3\%$ at the recommended SiPM operating voltage. At a light yield of 10 photoelectrons (p.e.), the expected dark count rate of the SiPMs is 3\,$\mu$Hz when modeled by the Borel distribution~\cite{Borel}. 




\paragraph{SiPM PCBs, mounts, and  coupling}

\begin{figure}[ht!]
\centerline{\includegraphics[scale=.25] {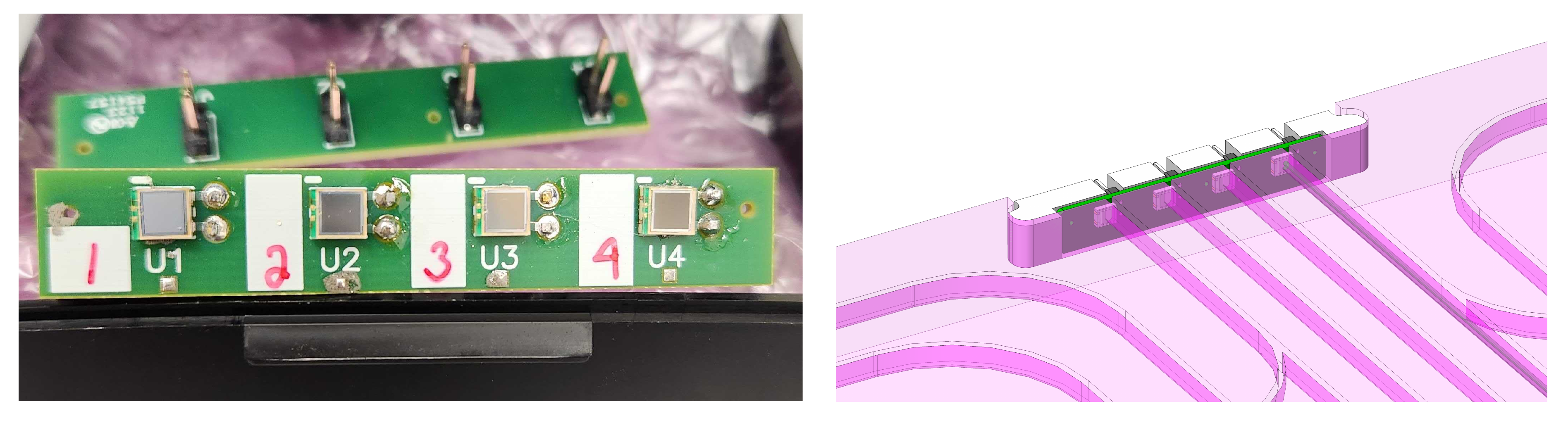}}
\caption{\label{SiPMMounts} Custom SiPM PCBs (left) and panel mounts (right) used in Prototypes 1 and 2.}
\end{figure}

Custom printed circuit boards (PCBs), shown in figure~\ref{SiPMMounts}, were designed to hold 4 SiPMs each for Prototypes 1 and 2. A ribbon cable supplies power to each SiPM and transmits its signal to the data acquisition (DAQ) system as described in section \ref{subsec:readout}. The boards are installed inside cutouts in the plastic scintillator using custom 3D-printed plastic holders. A small layer of Eljen Technologies EJ-550 silicon optical grease wass applied to the SiPMs to couple them to the ends of the WLS fibers~\cite{eljenEJ550}.

The Small Panel uses one 13360-3050 Hamamatsu SiPM soldered to an Adafruit Perma-Proto Small Tin board~\cite{Adafruit-board} and mounted to the Small Panel with a custom 3D-printed holder. The WLS fibers protrude $\sim$1\,cm from the panel, as shown in figure~\ref{SmallPanel}, and are held nearly flush against the SiPM by the mounted holder. 

\paragraph{SiPM testing}
Following PCB fabrication, every SiPM underwent the following quality assurance (QA) tests to ensure there was no damage during soldering and assembly:
\begin{itemize}
    \item Cable continuity checks for each ribbon cable connection.
    \item Visual inspection with a microscope to ensure SiPMs were not damaged during assembly.
    \item Waveform inspection using an oscilloscope, external power supply, and CAEN SP5601 LED Driver~\cite{LED-Driver} .
    \item Single photoelectron spectrum collected using the DAQ and LED driver.
\end{itemize}

\begin{figure}
\centerline{\includegraphics[scale=.5] {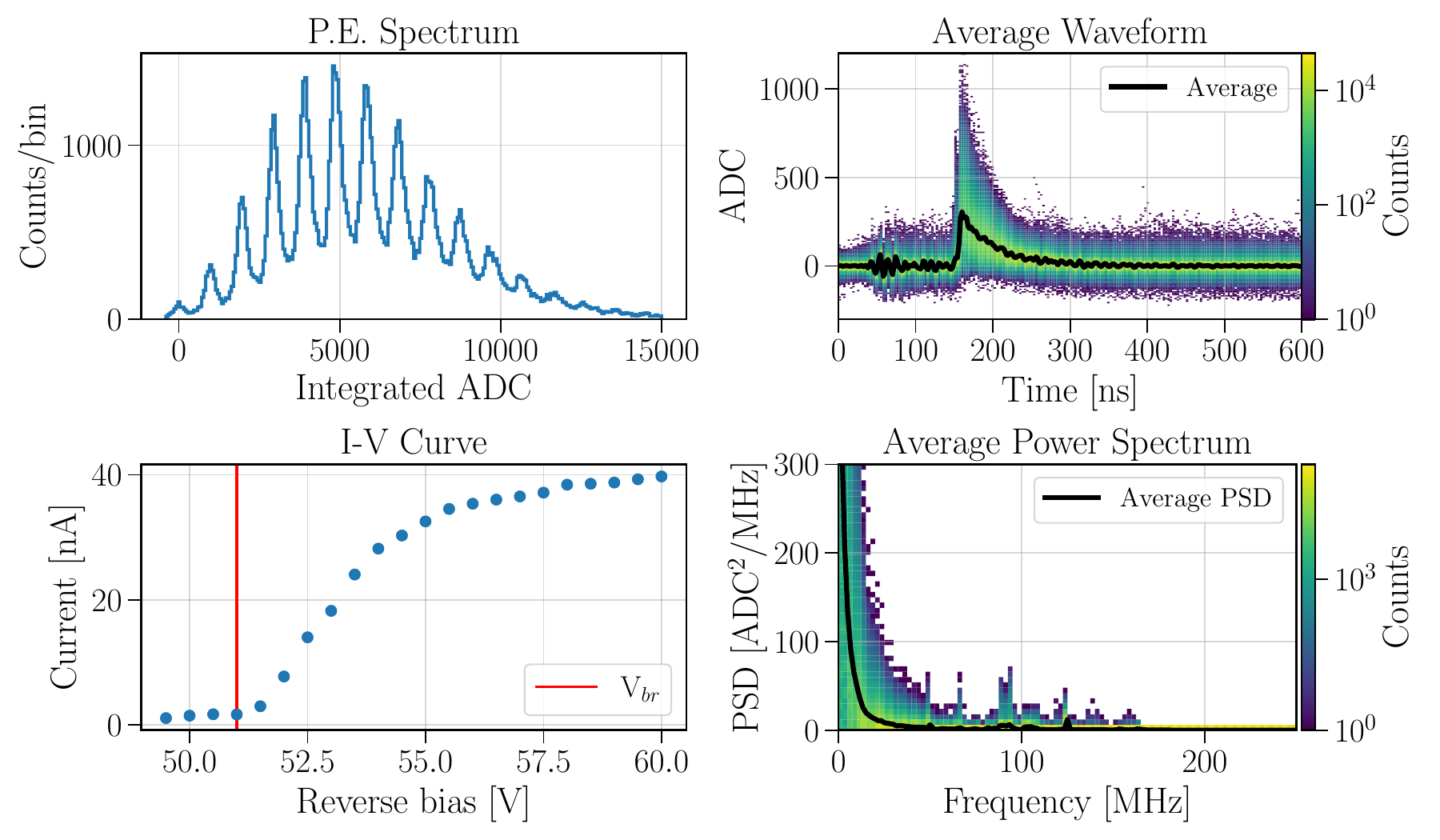}}
\caption{\label{SiPMQA} Tests in the developed SiPM Quality Assurance procedure. These plots are shown for a SiPM that passed our QA tests. It was biased at 55\,V (V\textsubscript{br} $\sim$ 51\,V) and flashed with an LED driver which had a wavelength of 405\,nm and amplitude dial set to 450~\cite{LED-Driver}. Waveforms were collected using a CAEN V1730 digitizer~\cite{digitizer}. The photoelectron (p.e.) spectrum of the SiPM is found by integrating 95,000 baseline-subtracted waveform samples over a window of [70,120] units of ADC, corresponding to a time window of [140, 240]\,ns (top left). These waveform samples are plotted and averaged to obtain the average waveform for the SiPM. The visible "ripple" around 100\,ns in the average waveform is the result of some electronic noise in the oscilloscope setup (top right). The current-versus-voltage (I-V) curve for the SiPM is measured with a Keithley 6487 Picoammeter~\cite{keithley}, with the breakdown voltage (V\textsubscript{br}) indicated in red; error bars are present but not visible due to the 10\textsuperscript{-5}\,nA precision of the picoammeter (bottom left). Finally, the power spectral density (PSD) is plotted for each waveform sample--the same as those shown in the top right--and used to get the average PSD for the SiPM (bottom right).}
\end{figure}

The SiPMs were operated at 55\,V (breakdown voltage V\textsubscript{br} $\sim$ 51\,V) during all data collection, while the overvoltage was modified during QA to minimize noise and dark count rates. The light intensity of the LED driver was set to a value of 450 (arbitrary units) via the amplitude dial, and has a typical wavelength of 405\,nm. The results of these tests are shown in figure~\ref{SiPMQA} for a well-functioning SiPM. Currently, our setup suffers from the addition of noise due to inadequate grounding and electronic interference, as can be seen in figure~\ref{SiPMQA}. A dedicated investigation to mitigate this noise is ongoing.

\paragraph{Future QA development}
A more rigorous QA procedure is being developed for use in the full muon veto system and has been performed on a sample of the SiPMs. In addition to the cable continuity and visual inspection tests described in the section above, this procedure includes measuring SiPM breakdown voltages after soldering and uses the CAEN V1730 digitizer~\cite{digitizer} to obtain waveforms (our current readout does not contain a digitizer, cf. section~\ref{subsec:readout}), Fourier transforms of those waveforms, and single photoelectron spectra.

\subsection{Electronics, readout, and DAQ}
\label{subsec:readout}

A CAEN DT5202 readout module was chosen for the muon veto data acquisition (DAQ) system~\cite{CAEN_DAQ}, having single photoelectron energy resolution and a timing resolution of 0.5\,ns. This unit is designed to provide a bias voltage to SiPMs and read out the signals of a large array of detectors using a single cable. CAEN's DT5202 readout software was modified to store the output in \texttt{binary} files for posterior use by the analysis framework \texttt{Janus}~\cite{JANUS}. Information such as integrated charge, trigger time, and timestamp is stored in the output file. The CAEN DT5202 module contains two Citiroc-1A chips~\cite{Citiroc}, each consisting of 32 channels, for a total of 64 channels. The relevant trigger modes are described below in brief; please consult the CAEN DT5202 manual for further detail~\cite{CAEN_DAQ}.

The discriminator output for each the module's 64 channels acts as a self-trigger, and the \texttt{TLOGIC} trigger source generates a trigger signal from logical combinations of these self-triggers. For Prototype 1, the 2.5\,cm single panel module, the \texttt{T-OR} logic trigger is used. In this mode, an event is triggered when any of the active CAEN DT5202 channels self-trigger, and the output signals from all active channels are saved. The \texttt{AND2\_OR32} logic trigger is used for Prototype 2, which consists of a pair of 1\,cm panels operating in coincidence. In this mode, an event is triggered whenever any pair of consecutive (even, odd)-ordered channels (e.g., channels 0 and 1, or channels 2 and 3, but not channels 1 and 2) from a Citiroc-1A chip have self-triggers that overlap by at least one clock cycle, or 8\,ns. The panel on the top in Prototype 2 is connected even channels (0, 2, ..., 14) and the bottom panel is connected to odd channels (1, 3, ..., 15).

\begin{figure}
\centerline{\includegraphics[scale=.6] {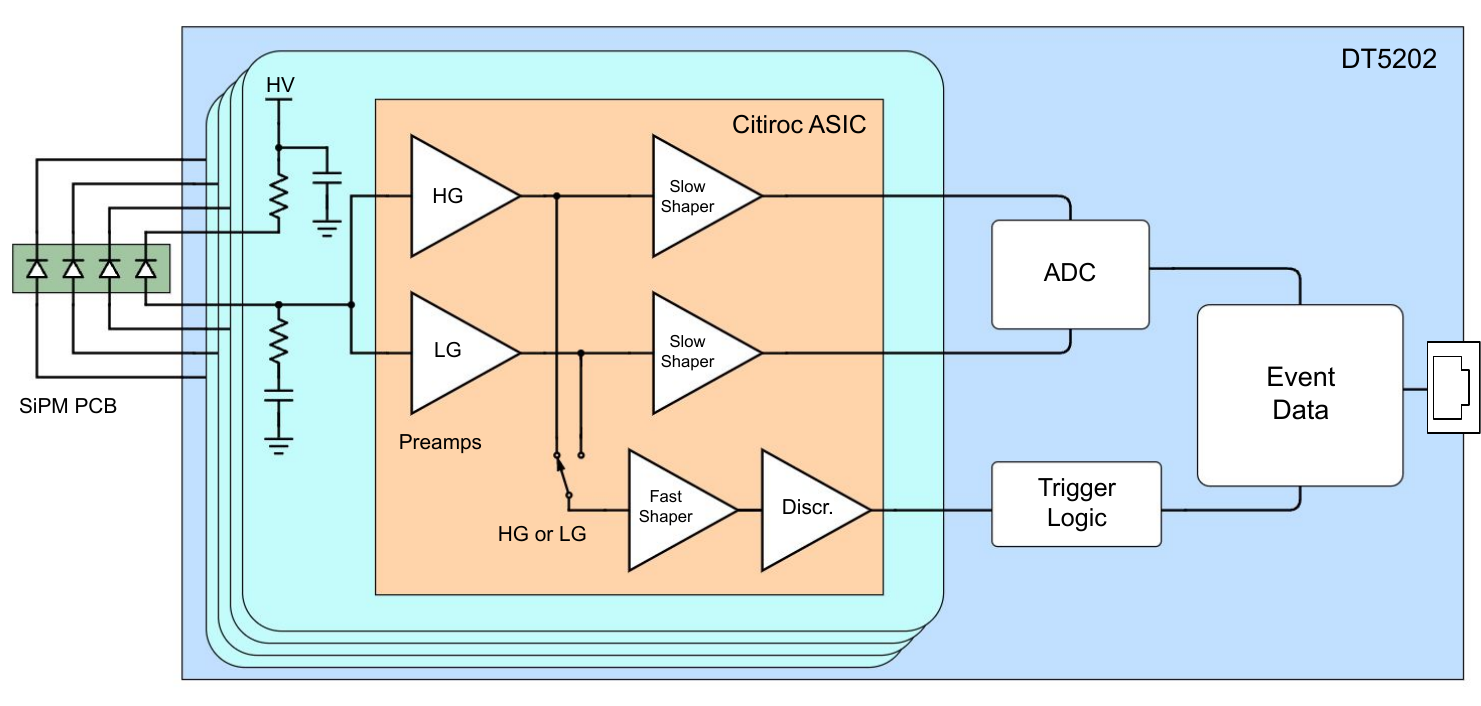}}
\caption{\label{daq_schem} Block diagram showing the relevant DAQ components and SiPM connection. Each rounded cyan rectangle represents a circuit connected to a single SiPM. The high gain (HG) and low gain (LG) preamplifier outputs are simultaneously processed by their own slow shapers, and triggering is done on either the HG or LG output. The ADC receives a multiplexed signal output from the Citiroc chip~\cite{Citiroc}. The PC receives output from the CAEN DT5202 via ethernet.}
\end{figure}

Input signals received by the CAEN DT5202 are sent simultaneously through two pre-amplifiers: a high gain amplifier and a low gain amplifier (see figure~\ref{daq_schem}). There is approximately a  $10\, \times \,$ difference in amplification between the two. This functionality allows for access to both single p.e. peaks and the broader energy spectrum for a single dataset via the high gain and low gain information, respectively.

Self-trigger threshold values were determined from staircase scans over a range of threshold values performed with the Janus software. Staircase scans were completed with two different setups: SiPMs decoupled from the panel placed inside a dark box with a reverse bias voltage, and SiPMs coupled to the panels with no bias voltage. From the first scan, a threshold value is chosen to minimize the event rate attributed to the intrinsic dark current at environmental temperature. The second scan helps to verify the DAQ is not triggering on electrical noise. The higher of the threshold values was set for all future runs taken with the panels.

\subsection{Data taking}

\begin{figure}[h!]
\centerline{\includegraphics[scale=.25] {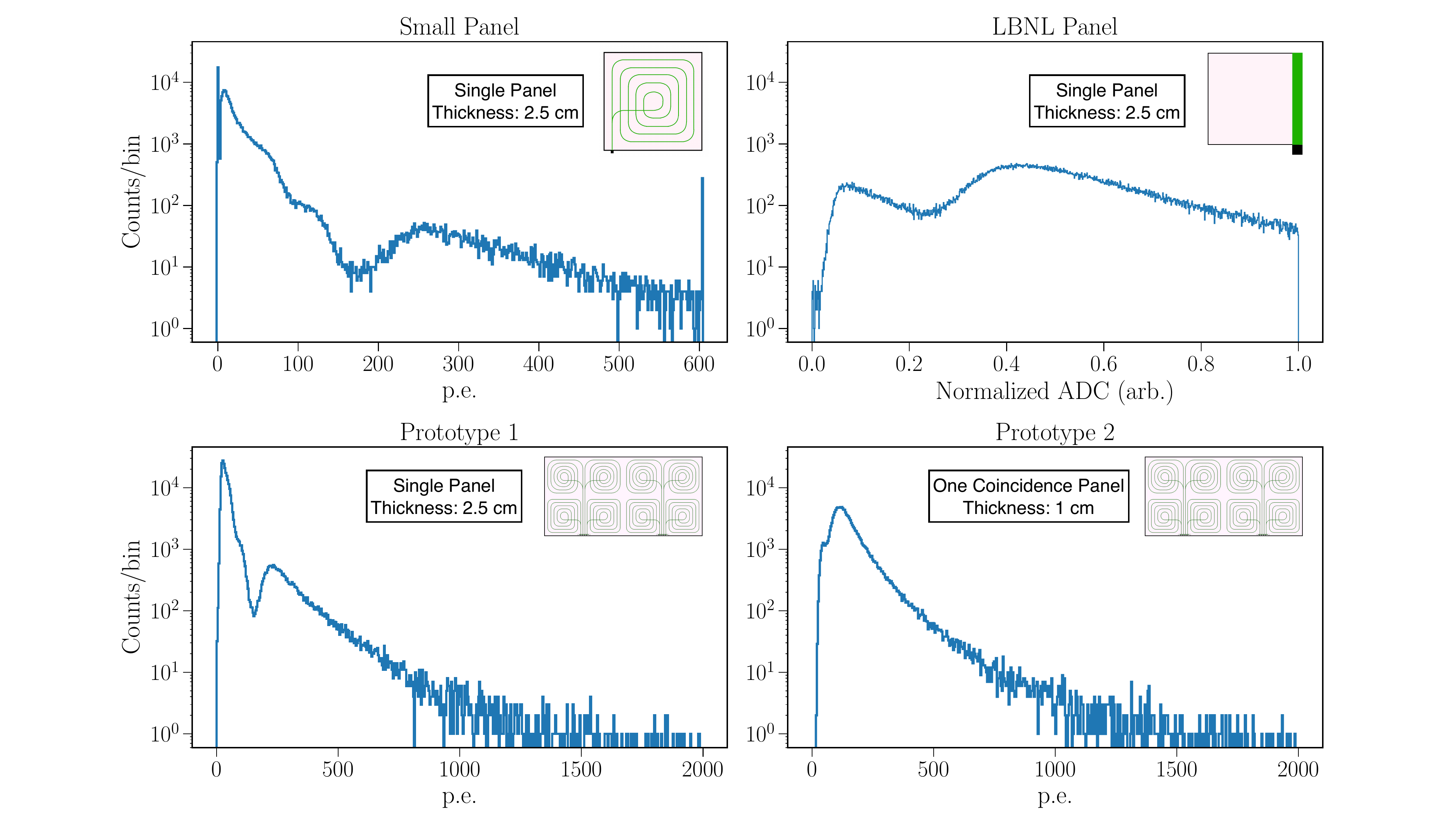}}
\caption{\label{MVBackgroundSpec} Background spectrum at about sea level for the Small Panel (top left), the LBNL panel (top right), Prototype 1 (bottom left), and one of the Prototype 2 panels (bottom right). The Prototype 1 and 2 panel spectra are the sum of 8 readout channels; the LBNL panel was not energy calibrated as we were only interested in its light uniformity. Details for the different prototypes can be found in section~\ref{MV_prototypes}.}
\end{figure}

Data collection for these prototypes occurred at the Yale Wright Laboratory in New Haven, CT, which is 22.8\,m above sea level. Panel prototypes were placed in custom dark boxes and covered with blackout fabric to reduce noise from ambient light in the SiPMs, see the resulting background spectra for each prototype in figure~\ref{MVBackgroundSpec}. This light tightness will be a requirement for the official installation at LNGS as artificial illumination will be present onsite. SiPMs were connected to the DAQ via CAEN A5261 remotization cables~\cite{caen-cables} and AMPMODU type 3-102203-3 connectors~\cite{cable-headers}.

\section{Panel performance}

\subsection{Light yield}
\label{subsec:LY}
The light yield (LY), described in units of p.e./MeV, is defined as
\begin{equation} \label{LYeq}
    \text{LY} = \frac{\text{ADC}_{\text{LG}}}{\text{MeV}} \, \times \, \frac{\text{p.e.}}{\text{ADC}_{\text{HG}}} \, \times \, \frac{\text{ADC}_\text{HG}}{\text{ADC}_{\text{LG}}}
\end{equation}
where ADC\textsubscript{LG} and ADC\textsubscript{HG} refer to the DAQ low gain and high gain readout, respectively (see section~\ref{subsec:readout}). The high gain and low gain are both needed for this calculation as the muon peak is visible only in the low gain (the relatively high energies of muon events get saturated in the high gain) and the high gain amplification is required to distinguish individual p.e. peaks.

\begin{figure}[H]
\centerline{\includegraphics[scale=.47] {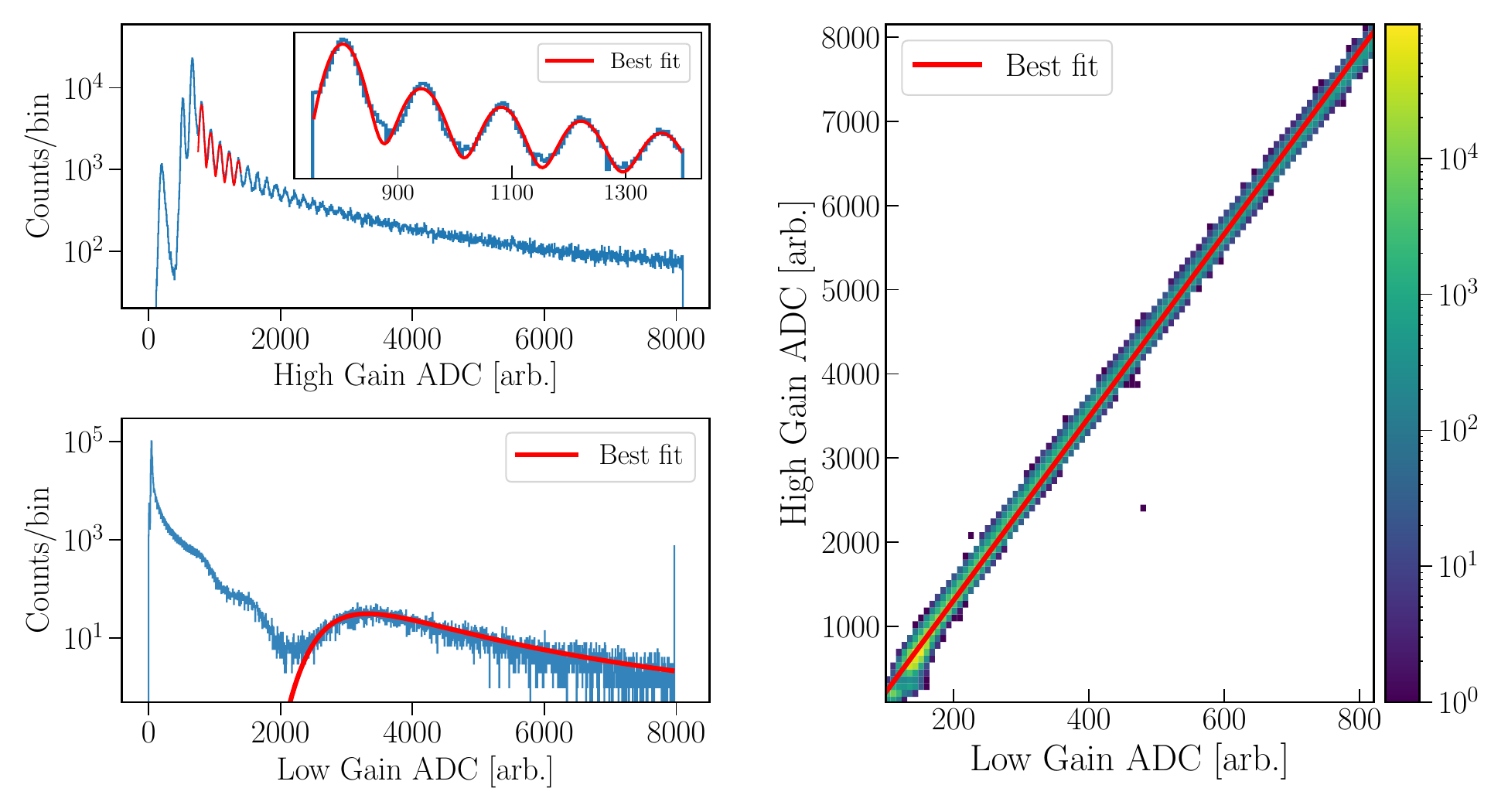}}
\caption{\label{swirlSpect} The Small Panel light yield was calculated using the spectra and fits shown here. Distinct single p.e. peaks are visible up to $\sim$2500 ADC in the high gain spectrum, a portion of which are fit with a sum of Gaussians to calculate the high gain ADC per p.e. ratio. The inset shows just the range of the high gain spectrum used in the fit (top left). The muon peak in the low gain spectrum is fit with a Landau convoluted with a Gaussian; the MPV of this fit corresponds to the low gain ADC per average muon energy deposition of 1.96\,MeV/cm (bottom left). The rightmost bin contains very high energy events (an overflow bin is also present in the high gain spectrum but is not shown in the figure for sake of readability). Finally, a 2D histogram plotting the high and low gain spectra is used to determine the conversion factor between high and low gain ADC (right).}
\end{figure}

The first term in equation~\ref{LYeq} comes from the most probable value (MPV) from the Landau-Gaussian convolution fit to the muon peak in the low gain data (see bottom left of figure~\ref{swirlSpect}). The MPV is divided by the average energy deposited in a panel by a muon. This value was estimated by simulating our setup in \texttt{Geant4}~\cite{Geant4, Geant4_2} using a $\cos^2(\theta)$ flux distribution, which yielded a value of ($4.98\pm0.02$)\,MeV for a 2.5\,cm thick panel. 

The second term is obtained by fitting the p.e.\ peaks in the high gain data with a series of independent Gaussians and averaging the difference in ADC between adjacent peaks. The fit is performed over a portion of the full ADC\textsubscript{HG} range that contains distinct peaks with relatively similar amplitudes (see top left of figure~\ref{swirlSpect}). Variations in the difference between adjacent peaks arise from statistical fluctuations and are consistent across the full ADC\textsubscript{HG} range. The independence of these fluctuations was confirmed by performing fits of the full range of visible p.e. peaks and various sub-ranges. The fit shown in figure~\ref{swirlSpect}, for example, includes five p.e. peaks over the range [750, 1400]\,ADC\textsubscript{HG} and yields a value of $140.0\,\pm \,1.7$\,ADC\textsubscript{HG}/p.e. Extending the fit to include the first 11 p.e. peaks over the range [470, 1990]\,ADC\textsubscript{HG} (this does not include the first peak at $\sim$0\,ADC\textsubscript{HG} which corresponds to 0\,p.e. (pedestal) events) yields a value of $141.0\,\pm\,3.2$\,ADC\textsubscript{HG}/p.e., which is consistent within uncertainty. 

Finally, the third term represents the ratio between the high gain and low gain ADC from calibration data. This comes from plotting the high gain data against the low gain data and performing a linear fit. Only events that do not saturate the DAQ's high gain mode are used for this calculation, and the linear trend is extrapolated to higher energy events (see figure~\ref{swirlSpect}). 

 To account for non-homogeneity between mini-modules in Prototype 1, the quantities in equation~\ref{LYeq} are calculated for each mini-module individually. The $\frac{\text{ADC}_{\text{LG}}}{\text{MeV}}$ values of each readout channel are summed and the two latter terms are determined by a weighted average based on each mini-module's contribution to the total $\frac{\text{ADC}_{\text{LG}}}{\text{MeV}}$ of the panel. 
The homogeneity of the mini-modules themselves was confirmed by operating the Small Panel in coincidence with a 25$\, \times \,$6$\, \times \,$2.5\,cm\textsuperscript{3} piece of plastic scintillator with a single WLS fiber and SiPM. The performance of the Small Panel was consistent for muon events in each region, indicating that the mini-module design provides sufficient homogeneity.
We found a light yield of ($52.9\,\pm \,0.9$)\,p.e./MeV for the Small Panel and of ($55.9 \,\pm \,1.5$)\,p.e./MeV for the full Prototype 1 panel. The light yield was not calculated for the Prototype 2 panels because the 1\,cm thickness does not produce a separation between the gamma and muon peaks in the ADC\textsubscript{LG} spectrum that is sufficient for the calculation to be reliable.

\subsection{Light collection uniformity}
\begin{figure}[h]
\centerline{\includegraphics[scale=.65] {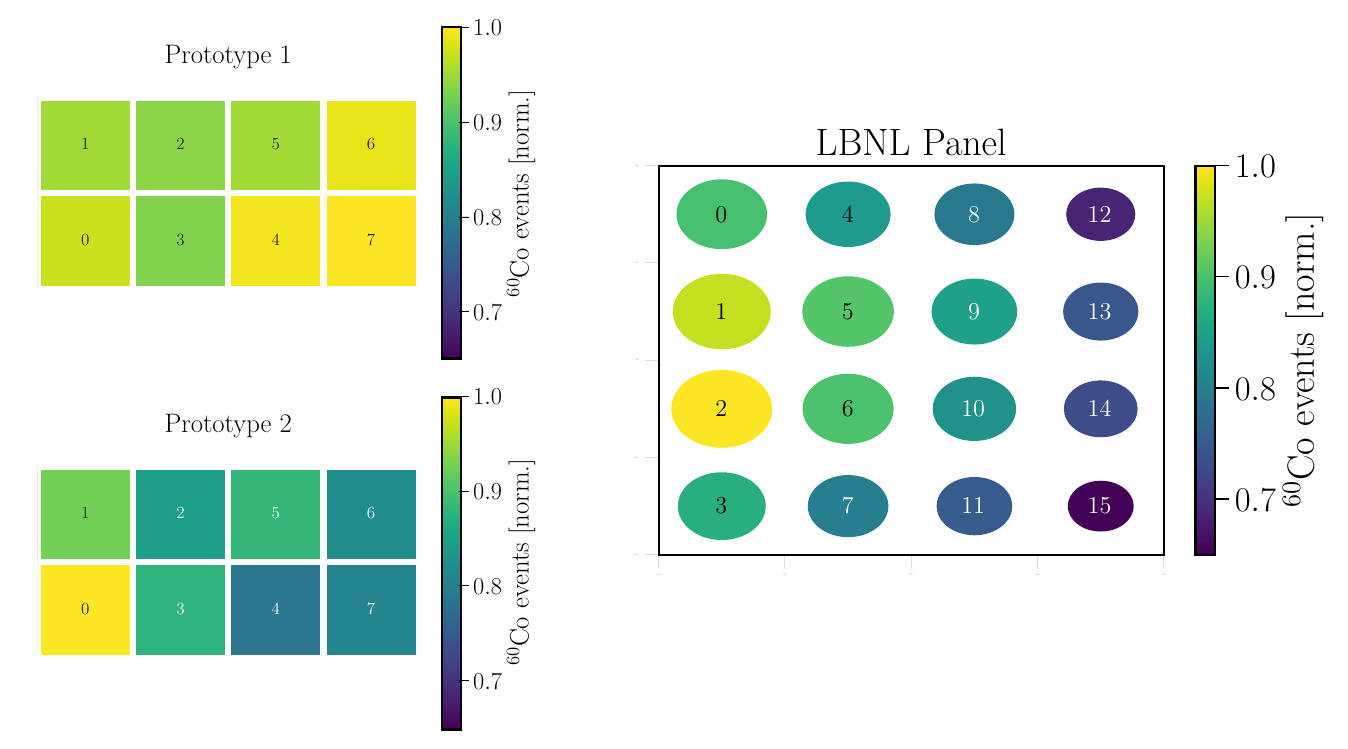}}
\caption{\label{MVLightCollectionHeatMap} Heat map of the normalized light collection for each mini-module of the Prototype 1 panel (top left), a Prototype 2 panel (bottom left), and for each $^{60}$Co source location for the LBNL Panel (right). Each mini-module is labeled from 0-7 for Prototype 1 and 2, which are oriented such that the SiPMs are located along the bottom row of the heat map. The LBNL Panel has no physical segmentation, but the $^{60}$Co source locations are labeled from 0-15. The LBNL Panel heat map is oriented as shown in figure~\ref{MV_prototypes}, with the WLS strip on the left side and the PMT in the top left corner. The color scale reflects the light collection of each location relative to the highest light-collecting location.}
\end{figure}

To characterize the light collection uniformity of the LBNL Panel, Prototype 1, and Prototype 2, a collimated $^{60}$Co source is placed at the center of each mini-module of the Prototype 1 and 2 panels, and at 16 uniformly-spaced positions across the LBNL Panel. For Prototype 2, a single 1\,cm thick panel was characterized. Data was taken for 5 minutes at each position of the source for the LBNL Panel and for 2 minutes for Prototypes 1 and 2. The ADC\textsubscript{LG} output from each source location is normalized to that of the source location that produced the highest number of counts in a given module, allowing for comparison between prototypes. The variations in light collection for each location of the $^{60}$Co source are shown in figure~\ref{MVLightCollectionHeatMap}.

To quantify the uniformity of the different panels, the total ADC\textsubscript{LG} spectrum is integrated for each data-taking interval (i.e., for each source location the total output from the panel was summed). The summation represents the total number of events observed by all mini-modules in the data-taking interval. The LBNL Panel is not physically segmented, so the summation represents the number of events detected by the PMT for the whole panel. Although this summation includes muon and gamma events in addition to the $^{60}$Co events, we can assume their respective fluxes to be uniform over time and to not represent a source of bias in our analysis. As described above, the source location with the largest sum of ADC\textsubscript{LG} counts is used to normalize all other locations. The statistical error is less than 1\% for all data points. Prototype 1 has the most uniform light collection. The average normalized number of $^{60}$Co events is 0.97\,$\pm$ 0.02, where the error is the statistical uncertainty. For Prototype 2, the uniformity is slightly worse with an average of 0.87\,$\pm$ 0.06 normalized $^{60}$Co events. The least uniform light collection was observed in the LBNL Panel with an average of 0.83\,$\pm$ 0.09 normalized $^{60}$Co events.

The behavior of light collection changes in a structured way for the LBNL Panel due to the unsegmented design. The light collection is highest for locations closest to the PMT and the light collection worsens for locations along the edge of the panel. The motivation for the swirly design of the fibers in Prototypes 1 and 2 was to increase the light collection uniformity, which is what we observe. This allows these panels to have higher efficiency in tagging muons. However, this introduces a source of systematics that affects our panels uniquely: the alignment of the WLS fibers with the SiPMs. Although we tried to minimize this effect by using a stencil during fabrication, the small area of the SiPM made this challenging. Since the diameter of each of the two fibers are 1\,mm, and they feed into a 3$\, \times \,$3\,mm$^2$ SiPM, there is little room for error in their alignment. This variation can contribute to the light collection fluctuation within a panel, as well as across panels.

\subsection{Muon discrimination power}
\label{sec:muonDiscPower}
\begin{figure}[h]
\centerline{\includegraphics[scale=.6] {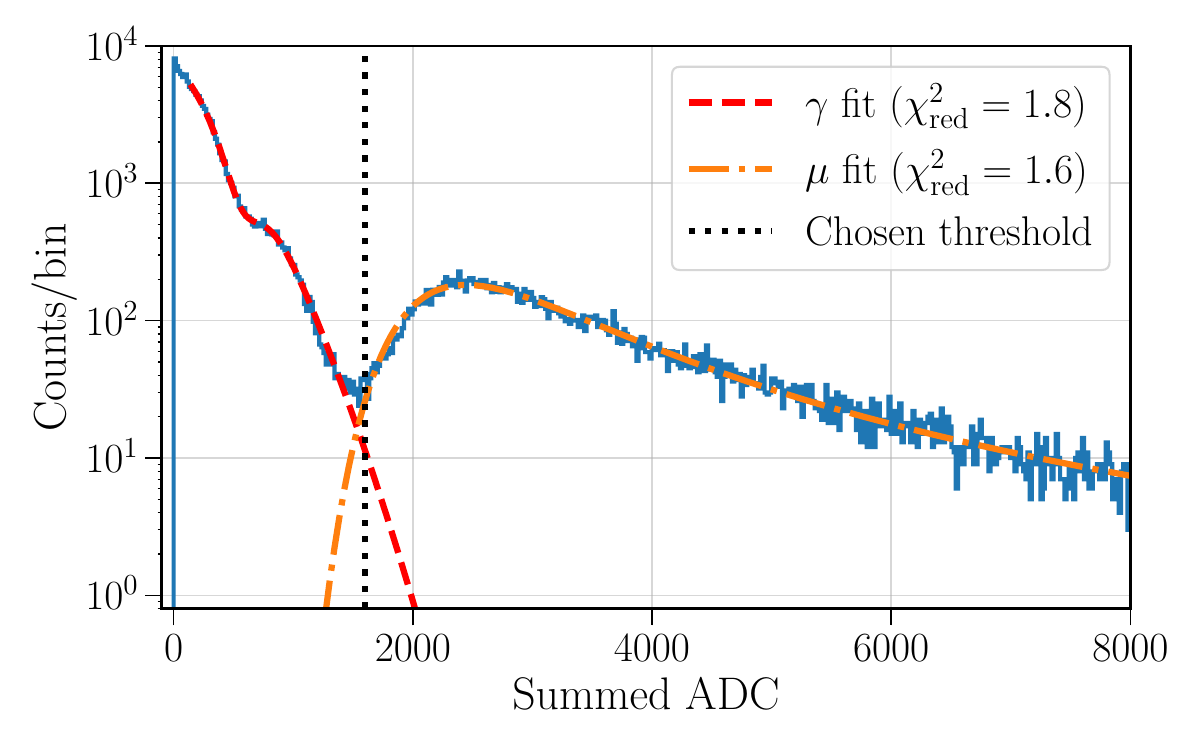}}
\caption{\label{muonDiscr} Energy spectrum for Prototype 1 with gamma peaks fit to a sum of Gaussians, muon peak fitted with a Landau-Gaussian distribution, and the ADC threshold indicated with a black dotted line. This energy spectrum was used to estimate the false-positive rate and muon discrimination efficiency described in section~\ref{sec:muonDiscPower}.}
\end{figure}

Muon discrimination refers to the ability to differentiate between muon and gamma events, whose distributions overlap in our panels' energy spectra. For a given (low gain) ADC threshold, events with a greater ADC value (to the right of the threshold) are tagged as muons. We estimate the fraction of all muons that get tagged using a data-driven approach in which the gamma peaks follow a Gaussian distribution whereas the muon events are distributed according to a Landau-Gaussian distribution, as shown in figure~\ref{muonDiscr}. 

The muon tagging efficiency is obtained by integrating the Landau-Gaussian distribution from a chosen ADC threshold, which separates the two distributions, and normalizing by the marginalized Landau-Gaussian distribution. We chose this threshold to be 1600\,ADC. We find that the ratio of muons that are tagged above that threshold for Prototype 1 is $(99.3\,
\substack{+0.7\\-1.0})\%$. Similarly, for that particular threshold choice, the false positive rate from gamma events corresponds to 0.34\,Hz for a self-triggered single module. To estimate the induced deadtime, we conservatively~\cite{Haffke_2011,JRC135812} assume a similar false-positive rate underground at LNGS, and an estimated coverage area of 78\,m$^2$ by the complete muon veto system. We expect an induced deadtime of $<1\%$ in CUPID for a veto window of 0.5\,ms (CUPID's light detector rise-time~\cite{CUPID:2025avs}) by assuming coincidence hits between two panels intersected by the muon track. This meets CUPID's deadtime requirement of $<1\%$, and is mainly dominated by false-positive events from the gamma peak leaking into the muon peak, making the induced rate from real muon events negligible. This estimation was not performed for Prototype 2 since the gamma/muon separation is not resolvable.

\begin{figure}[ht!]
\centerline{\includegraphics[scale=.25] {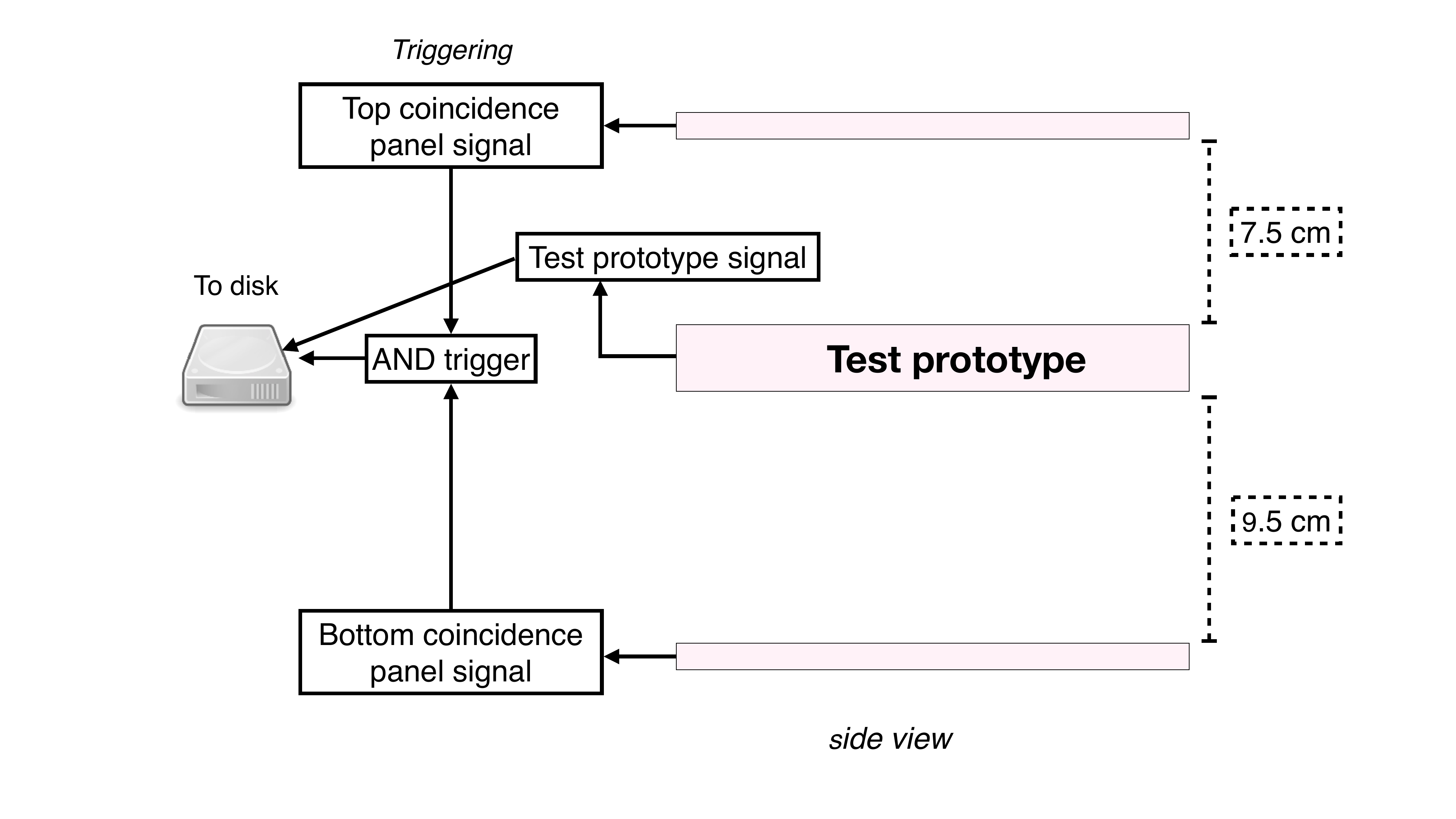}}
\caption{\label{fig:PanelEff} Setup to estimate the detection efficiency of the single module. The thin, outer scintillator panels trigger in coincidence when a muon passes through both of them. The test prototype in the middle records data upon a trigger but does not participate in the triggering scheme.}
\end{figure}


\subsection{Detection efficiency}
\label{subsec:deteff}
We estimate the detection efficiency, i.e. the fraction of muons detected (above a chosen threshold, see Fig.~\ref{muonDiscr}) compared to the total number of muons that pass through the single module. To do this, we "sandwich" the Prototype 1 module between the two Prototype 2 panels, which trigger in coincidence, while the channels in the Prototype 1 module are masked out and do not participate in the triggering scheme. The panels are physically separated as shown in figure~\ref{fig:PanelEff}, with an arbitrarily chosen separation length determined mainly by the dimensional constraints of our dark box.

With this configuration, we estimate the detection efficiency of Prototype 1 to be $\epsilon = (0.98\pm0.01)$. We assume a detection efficiency of $\sim 1$ for Prototype 2, so we do not apply any corrections to the final number that we report here. Although this efficiency is already high and consistent with other similar designs~\cite{SEO2022167123Amore}, it can be further improved by increasing the thickness of the scintillator panel or operating in coincidence using two stacked modules instead of one. This might be necessary for ultra-low background experiments with stringent background budgets.

\subsection{Position reconstruction of muon events}
One motivation for including eight readouts in each panel of Prototypes 1 and 2 was to enable coarse position reconstruction of events. To quantify the position reconstruction of Prototype 1 for muon events, eight 30-minute runs were taken using the Small Panel prototype (25$\, \times \,$25\,cm$^2$ panel) in a coincidence configuration with the Prototype 1 mini-modules, following the same location labeling as figure~\ref{MVLightCollectionHeatMap}. \texttt{OR} trigger logic was used for data taking, and coincidence events were found offline between the Small Panel and Prototype 1. In this analysis, muon events were considered to be coincidence events above a defined (low gain) ADC threshold in the Small Panel's energy spectrum. Because the Small Panel rests directly on top of Prototype 1, the position of a muon event is assumed to be in the same area as the location of the Small Panel for each data run.


\begin{figure}
\centerline{\includegraphics[scale=.55] {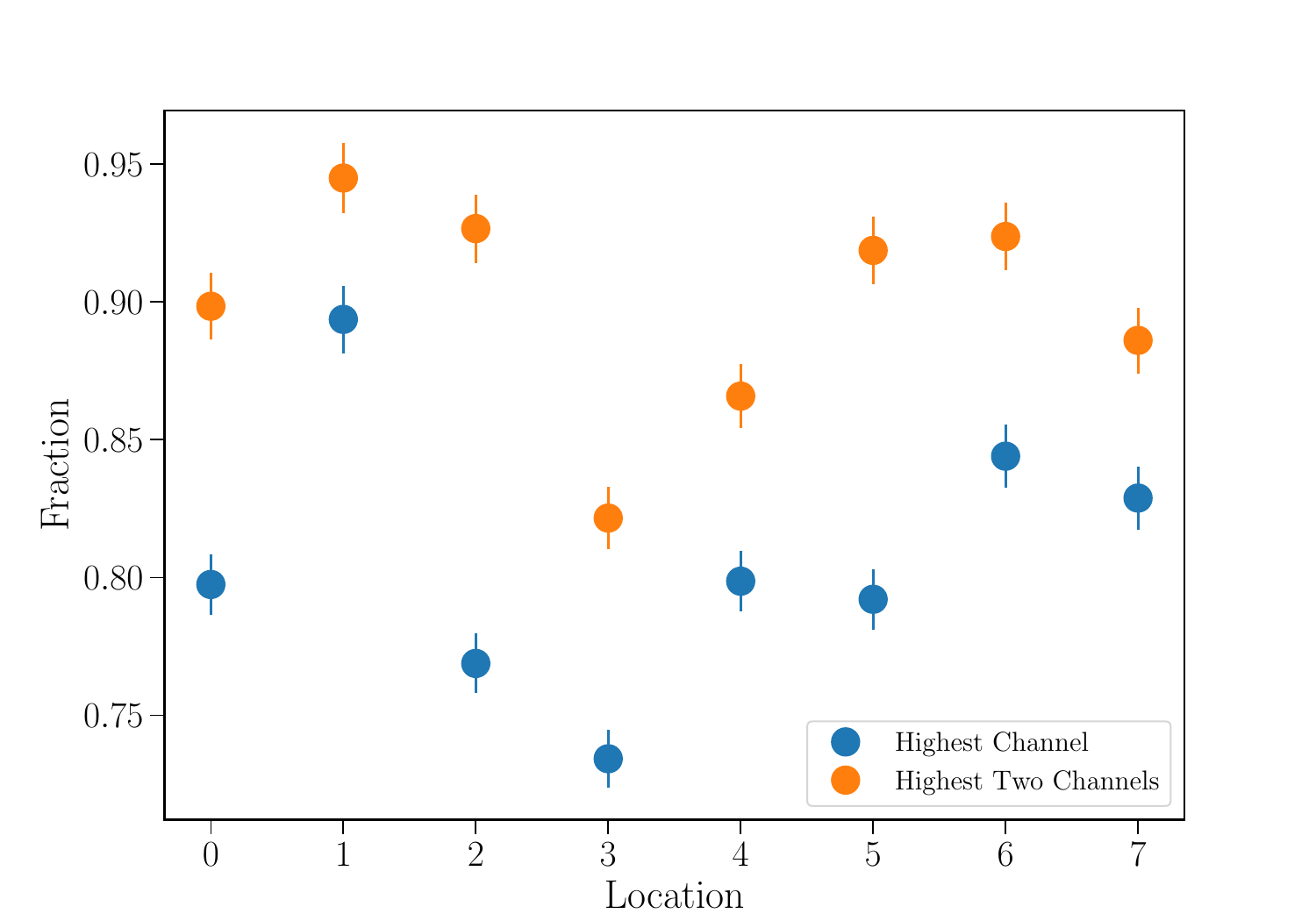}}
\caption{\label{CoRunPositionsReconstruction} The fraction of muon events in Prototype 1 whose position can be identified by the SiPM channel with the highest light yield or the two highest SiPM channels based on location. For this study, one hour of background data was collected with Prototype 1 and the 25$\, \times \,$25\,cm$^2$ Small Panel as a coincidence panel at different locations. Muon events are considered to be coincident events above an energy threshold defined according to the Small Panel spectrum. The location of the muons is assumed to be in the 25$\, \times \,$25\,cm$^2$ area of coincidence.}
\end{figure}

For each location, the spectrum (in p.e.) from each of the Prototype 1 SiPM channels, the Small Panel, and the sum of Prototype 1 channels were examined. In most locations, the SiPM channel with the highest muon peak p.e. in Prototype 1 corresponds to the mini-module directly under the Small Panel. This suggests that the sub-section of the panel the muon passes through typically has the highest light yield. With these observations, two simple algorithms were designed to identify the location of muon events and estimate position reconstruction capabilities. The first algorithm assigns the location of a muon event to the mini-module channel with the highest muon peak p.e., which covers a 25$\, \times \,$25\,cm\textsuperscript{2} area. The second algorithm assigns both the mini-module channel with the highest and the channel with the second highest muon p.e. peak to each muon event, reconstructing the event to a 25$\, \times \,$50\,cm$^2$ area. These algorithms were run on the eight collected datasets for the Prototype 1 module (before any coincidence cuts) to test their efficiencies as the muon events in each location can be isolated and are known. The fraction of correctly reconstructed muon events is shown in figure~\ref{CoRunPositionsReconstruction}. The track reconstruction algorithm using the highest two channels had an efficiency of $(89.8 \,\pm \,0.4)\%$, and the single track channel reconstruction had an efficiency of $(80.7 \,\pm \,0.4)\%$.

Additional position reconstruction algorithms could be developed for track reconstruction using more SiPM channels and metrics within a panel such as the ratio of ADC counts in an event, multiple panels, and information from events in CUORE crystals. This development is beyond the scope of this work. However, from the simple algorithms described, we estimate that using Prototype 1, the position of muon events is expected to reconstruct to 25$\, \times \,$25\,cm$^2$ with approximately an 80$\%$ efficiency. Because Prototype 2 operates as a double panel module between mini-modules of 25$\, \times \,$25\,cm$^2$, its position reconstruction granularity is assumed to be 25$\, \times \,$25\,cm$^2$ from the two adjacent channels that trigger in coincidence on the top and bottom panels. The possibility for finer position reconstruction is an advantage of including multiple mini-modules with their own light detectors in one larger module instead of using one light detector per module.

\section{Discussion}

Comparing the overall results of all prototypes, the swirly pattern of the WLS fiber used in the Small Panel and Prototypes 1 and 2 resulted in modules with high uniformity, light yield, and position reconstruction. Prototype 1 had the highest light yield and the capability to reconstruct the position of muon events to an area of 25$\, \times \,$25\,cm$^2$. Modules with a surface area of 100$\, \times \,$50\,cm$^2$ would be simpler to install and operate than 25$\, \times \,$25\,cm$^2$ panels. Additionally, modules with multiple SiPMs connected in parallel in the same continuous volume provide some contingency against broken SiPMs. Operating a Prototype 1 panel with only 7 SiPMs instead of 8 results in a detection efficiency of $\epsilon = (0.92\pm0.01)$, constituting a 6\% decrease from the fully operational panel (see section \ref{subsec:deteff}).

From the tested configuration (directly stacked 1\,cm panels), it was not evident that Prototype 2 would significantly decrease the gamma background, which would result in an increase of the detection efficiency and a decrease of the false-positive rate. Adding separation between two thin panels (Prototype 1), either air or another material such as lead, could potentially achieve this result. However, the cost would be an increase in the size of each module, which could conflict with CUPID's design constraints. Prototype 2 would also increase the number of channels by a factor of two. However, taking out the position requirement will further reduce the number of channels per panel.

While Prototype 1 has an adequate detection efficiency, an improved factor could be desired for ultra-low background experiments. A way to achieve this is to increase the thickness of the panels, yielding a greater gamma and muon peak separation and therefore a better discrimination power. These options for Prototypes 1 and 2 are currently being explored and optimized with the help of the information obtained from this study.

\section{Conclusions}

In this work, we designed, constructed, and characterized two novel prototypes for CUPID's compact muon veto system. The proposed prototypes consisting of wavelength-shifting fiber-embedded plastic scintillators coupled to silicon photomultipliers demonstrated a high and uniform light yield and a muon tagging efficiency of 98\%. These features will be critical to reduce muon-induced background and enable the CUPID experiment to reach a sensitivity level commensurate with its science goal. Additionally, these panels had efficient position reconstruction capabilities within an area of 25$\, \times \,$25\,cm$^2$. This is the first muon veto design that explores position reconstruction within a single panel using a swirly geometry. The position reconstruction capability could be useful for further decreasing backgrounds.

The prototypes in this paper were fabricated and characterized to study the performance of multiple single module designs. Several prototypes meet the muon veto requirements. Further modifications, such as reduced channels and the combination of wavelength-shifting fiber spirals, could be made to reduce the complexity of the system without having a considerable impact on the required metrics.

These prototypes successfully provided high efficiency in muon detection and position reconstruction. Future efforts will focus on optimizing a coincidence-based approach to further suppress the false-positive rate, and adapting the system for large-scale deployment in CUPID. Additional work still needs to be performed before the final CUPID muon veto design can be realized, including studies of prototype performance and longevity under the environmental conditions at the Laboratori Nazionali del Gran Sasso and consideration of any systematic effects that may arise from the muon distribution at Gran Sasso. In future work, we plan to address these uncertainties and the overall impact of all studies on the final CUPID muon veto design.


\acknowledgments
We thank the technical staff of Wright Laboratory, particularly Tom Barker for his assistance in designing PCBs. We acknowledge the contributions of Penny Slocum, James Nikkel, Thomas Barker, Caitlin Gainey, Gabriel Hoshino, Jacky Hua, Neal Ma, Din-Ammar Tolj, Iffat Zarif, and Andrew Zheng to this research. The authors are appreciative of the members of the CUORE and CUPID collaborations for their support and expertise in this work. This paper includes data from a panel produced for a muon veto at the Low Background Counting Facility at the Lawrence Berkeley National Laboratory. We thank Rick Norman for providing this panel and feedback on this manuscript. We also thank Vincenzo Caracciolo for useful discussions and comments on the manuscript. Eljen Technologies, particularly Chris Maxwell, provided samples and extensive knowledge for this project, for which we are grateful. Kuraray and Saint-Gobain (now named Luxium Solutions) provided fiber samples and expertise. We would also like to thank Hamamatsu Photonics and CAEN Technologies for their technical support, particularly Yuri Venturini and Marco Locatelli. This work is supported by the DOE Office of Science, Office of Nuclear Physics, Contract No. DE-SC0012654. R.N.G. acknowledges grant NSF-PHY-1913374. This work made use of \texttt{pandas}~\cite{reback2020pandas}, \texttt{Matplotlib}~\cite{Hunter:2007}, \texttt{Numpy}~\cite{harris2020array}, \texttt{SciPy}~\cite{2020SciPy-NMeth}, and \texttt{Jupyter}~\cite{Kluyver2016jupyter}.

\bibliographystyle{unsrt}
\bibliography{references}

\begin{thebibliography}{10}

\bibitem{PhysRevD.25.2951}
J.~Schechter and J.~W.~F. Valle.
\newblock {Neutrinoless double-$\ensuremath{\beta}$ decay in
  SU(2)\ifmmode\times\else\texttimes\fi{}U(1) theories}.
\newblock {\em Phys. Rev. D}, 25:2951--2954, Jun 1982.

\bibitem{NatureCUORE2022}
D.~Q. Adams et~al.
\newblock {Search for Majorana neutrinos exploiting millikelvin cryogenics with
  CUORE}.
\newblock {\em Nature}, 604(7904):53--58, 2022.

\bibitem{CUORE:2017tlq}
C.~Alduino et~al.
\newblock {First Results from CUORE: A Search for Lepton Number Violation via
  $0\nu\beta\beta$ Decay of $^{130}$Te}.
\newblock {\em Phys. Rev. Lett.}, 120(13):132501, 2018.

\bibitem{CUPID:2025avs}
K.~Alfonso et~al.
\newblock {CUPID, the CUORE Upgrade with Particle Identification}, 2025.
\newblock arXiv:2503.02894.

\bibitem{Cecchini_2009}
S.~Cecchini et~al.
\newblock Time variations in the deep underground muon flux.
\newblock {\em Europhysics Letters}, 87(3):39001, Aug 2009.

\bibitem{Bellini:2011yd}
G~Bellini et~al.
\newblock {Muon and cosmogenic neutron detection in Borexino}.
\newblock {\em Journal of Instrumentation}, 6(05):P05005, May 2011.

\bibitem{Mei:2005gm}
D.-M. Mei and A.~Hime.
\newblock Muon-induced background study for underground laboratories.
\newblock {\em Phys. Rev. D}, 73:053004, Mar 2006.

\bibitem{Rahaman:2007ng}
S.~Rahaman et~al.
\newblock {Q value of the Mo-100 Double-Beta Decay}.
\newblock {\em Phys. Lett. B}, 662:111--116, 2008.

\bibitem{Poda:2021}
D.~Poda.
\newblock {Scintillation in Low-Temperature Particle Detectors}.
\newblock {\em Physics}, 3(3):473--535, 2021.

\bibitem{CUORE:2024fak}
D.~Q. Adams et~al.
\newblock {Data-driven background model for the CUORE experiment}.
\newblock {\em Phys. Rev. D}, 110(5):052003, 2024.

\bibitem{GammaBackgroundLNGS}
C.~Bucci et~al.
\newblock {Background study and Monte Carlo simulations for large-mass
  bolometers}.
\newblock {\em The European Physical Journal A}, 41(2):155--168, 2009.

\bibitem{OMineev_2011}
O.~Mineev et~al.
\newblock Scintillator detectors with long wls fibers and multi-pixel
  photodiodes.
\newblock {\em Journal of Instrumentation}, 6(12):P12004, Dec 2011.

\bibitem{BUGG201491}
W.~Bugg, Yu. Efremenko, and S.~Vasilyev.
\newblock Large plastic scintillator panels with wls fiber readout:
  Optimization of components.
\newblock {\em Nuclear Instruments and Methods in Physics Research Section A:
  Accelerators, Spectrometers, Detectors and Associated Equipment}, 758:91--96,
  2014.

\bibitem{ZONG201882}
Z.~Zong et~al.
\newblock Study of light yield for different configurations of plastic
  scintillators and wavelength shifting fibers.
\newblock {\em Nuclear Instruments and Methods in Physics Research Section A:
  Accelerators, Spectrometers, Detectors and Associated Equipment}, 908:82--90,
  2018.

\bibitem{Luo:2023inu}
Guang Luo et~al.
\newblock {Design optimization of plastic scintillators with
  wavelength-shifting fibers and silicon photomultiplier readouts in the top
  veto tracker of the JUNO-TAO experiment}.
\newblock {\em Nucl. Sci. Tech.}, 34(7):99, 2023.

\bibitem{AHARONIAN2021165193}
F.~Aharonian et~al.
\newblock Performance test of the electromagnetic particle detectors for the
  lhaaso experiment.
\newblock {\em Nuclear Instruments and Methods in Physics Research Section A:
  Accelerators, Spectrometers, Detectors and Associated Equipment},
  1001:165193, 2021.

\bibitem{volchenko2008level}
V.~I. Volchenko et~al.
\newblock On the level of background in underground muon measurements with
  plastic scintillator counters, 2008.
\newblock arXiv:0810.2414.

\bibitem{SEO2022167123Amore}
J.W. Seo, E.J. Jeon, W.T. Kim, Y.D. Kim, H.Y. Lee, J.~Lee, M.H. Lee, P.B.
  Nyanda, and E.S. Yi.
\newblock A feasibility study of extruded plastic scintillator embedding wls
  fiber for {AMoRE-II} muon veto.
\newblock {\em Nuclear Instruments and Methods in Physics Research Section A:
  Accelerators, Spectrometers, Detectors and Associated Equipment},
  1039:167123, 2022.

\bibitem{Lu:2023sbi}
Peizhi Lu, Fengpeng An, Yu~Chen, Min Li, Yichen Li, Guang Luo, Wei Wang, Zhimin
  Wang, Xiang Xiao, and Y.~K. Hor.
\newblock {Study in the optical performance of plastic scintillator with WLS
  fiber}.
\newblock {\em JINST}, 18(04):T04002, 2023.

\bibitem{LBNLMuonVetoPanels}
K.J. Thomas, E.B. Norman, A.R. Smith, and Y.D. Chan.
\newblock {Installation of a muon veto for low background gamma spectroscopy at
  the LBNL low-background facility}.
\newblock {\em Nuclear Instruments and Methods in Physics Research Section A:
  Accelerators, Spectrometers, Detectors and Associated Equipment}, 724:47--53,
  2013.

\bibitem{eljenEJ200}
Eljen Technology.
\newblock {EJ-200 - Plastic Scintillator}.
\newblock
  \url{https://eljentechnology.com/products/plastic-scintillators/ej-200-ej-204-ej-208-ej-212}.
\newblock Accessed 25 Jan 2023.

\bibitem{Luxium}
Luxium Solutions.
\newblock Plastic scintillating fibers.
\newblock
  \url{https://www.luxiumsolutions.com/radiation-detection-scintillators/fibers}.
\newblock Accessed 19 Feb 2025.

\bibitem{kuraray}
Kuraray.
\newblock Plastic scintillating fibers (materials and structures).
\newblock \url{http://kuraraypsf.jp/psf/index.html}.
\newblock Accessed 7 Mar 2022.

\bibitem{Tyvek}
DuPont.
\newblock {DuPont}\texttrademark {T}yvek\textsuperscript{\textregistered}
  {8740D}.
\newblock
  \url{https://www.dupont.com/content/dam/dupont/amer/us/en/safety/public/documents/en/2019-C&I_Tyvek_8740D_Datasheet.pdf}.
\newblock Accessed 15 Jun 2025.

\bibitem{TyvekTape}
DuPont.
\newblock Tyvek\textsuperscript{\textregistered} {T}ape.
\newblock \url{https://www.dupont.com/products/tyvek-tape.html}.
\newblock Accessed 14 Jun 2023.

\bibitem{BlackTape}
Inc. Thorlabs.
\newblock High-performance black masking tape.
\newblock \url{https://www.thorlabs.com/thorproduct.cfm?partnumber=T743-2.0}.
\newblock Accessed 14 Jun 2023.

\bibitem{eljenMachiningPolishing}
Eljen Technology.
\newblock Machining and polishing of plastic scintillators.
\newblock
  \url{https://eljentechnology.com/images/technical_library/Machine_Polish_Plastics_2015.pdf}.
\newblock Accessed 25 Jan 2023.

\bibitem{NOA68}
Norland~Products Inc.
\newblock {Norland Optical Adhesive NOA 68}.
\newblock \url{https://norlandproducts.com/product/noa-68/}.
\newblock Accessed 10 Jun 2023.

\bibitem{eljenEJ500}
Eljen Technology.
\newblock Ej-500 - optical cement.
\newblock \url{https://eljentechnology.com/products/accessories/ej-500}.
\newblock Accessed 7 Mar 2022.

\bibitem{MPPCS13360-3050PE}
Hamamatsu Photonics.
\newblock {MPPC} {S13360-3050PE}.
\newblock
  \url{https://www.hamamatsu.com/eu/en/product/optical-sensors/mppc/mppc_mppc-array/S13360-3050PE.html}.
\newblock Accessed 13 Jun 2023.

\bibitem{StGobain_private}
Private correspondence with {Saint Gobain Crystals}, 2022.

\bibitem{eljenEJ560}
Eljen Technology.
\newblock Ej-560 - silicone rubber optical interface.
\newblock \url{https://eljentechnology.com/products/accessories/ej-560}.
\newblock Accessed 8 Jan 2024.

\bibitem{MPPCS13360-3050CS}
Hamamatsu Photonics.
\newblock {MPPC} {S13360-3050CS}.
\newblock
  \url{https://www.hamamatsu.com/eu/en/product/optical-sensors/mppc/mppc_mppc-array/S13360-3050CS.html}.
\newblock Accessed 13 Jun 2023.

\bibitem{Borel}
S.~Vinogradov.
\newblock Analytical models of probability distribution and excess noise factor
  of solid state photomultiplier signals with crosstalk.
\newblock {\em Nuclear Instruments and Methods in Physics Research Section A:
  Accelerators, Spectrometers, Detectors and Associated Equipment},
  695:247--251, 2012.
\newblock New Developments in Photodetection NDIP11.

\bibitem{eljenEJ550}
Eljen Technology.
\newblock Ej-550 - silicone optical grease.
\newblock \url{https://eljentechnology.com/products/accessories/ej-550-ej-552}.
\newblock Accessed 11 Jun 2025.

\bibitem{Adafruit-board}
Adafruit.
\newblock Perma-proto mint tin size breadboard {PCB}.
\newblock \url{https://www.adafruit.com/product/723}.
\newblock Accessed 8 Jan 2024.

\bibitem{LED-Driver}
CAEN.
\newblock {SP5601} {LED} {Driver}.
\newblock \url{https://www.caen.it/products/sp5601/}.
\newblock Accessed 13 Dec 2023.

\bibitem{digitizer}
CAEN.
\newblock {CAEN V1730} digitizer.
\newblock \url{https://www.caen.it/products/v1730/}.
\newblock Accessed 8 Jan 2024.

\bibitem{keithley}
Keithley.
\newblock {Keithley} {Model} {6497} {Picoammeter/Voltage Source}.
\newblock
  \url{https://www.tek.com/en/datasheet/series-6400-picoammeters/6487-picoammeter-voltage-source}.
\newblock Accessed 25 Feb 2025.

\bibitem{CAEN_DAQ}
{CAEN}.
\newblock {CAEN} {DT}5202 {DAQ}.
\newblock \url{https://www.caen.it/products/dt5202/}.
\newblock Accessed: 10 Mar 2022.

\bibitem{JANUS}
{CAEN}.
\newblock {JANUS} {FERS}-5200 {DAQ} software.
\newblock \url{https://www.caen.it/products/janus/}.
\newblock Accessed 14 Jun 2023.

\bibitem{Citiroc}
{CAEN}.
\newblock {Citiroc1A}.
\newblock \url{https://www.caen.it/products/citiroc-1a/}.
\newblock Accessed 25 Jan 2024.

\bibitem{caen-cables}
CAEN.
\newblock A5261 {SiPM} remotization cables.
\newblock \url{https://www.caen.it/products/a5261/}.
\newblock Accessed 8 Jan 2024.

\bibitem{cable-headers}
TE~Connectivity.
\newblock Te connectivity amp connectors 3-102203-4.
\newblock
  \url{https://www.digikey.com/en/products/detail/te-connectivity-amp-connectors/3-102203-4/299278}.
\newblock Accessed 8 Jan 2024.

\bibitem{Geant4}
J.~Allison et~al.
\newblock Geant4 developments and applications.
\newblock {\em IEEE Transactions on Nuclear Science}, 53(1):270--278, 2006.

\bibitem{Geant4_2}
J.~Allison et~al.
\newblock Recent developments in {Geant4}.
\newblock {\em Nuclear Instruments and Methods in Physics Research Section A:
  Accelerators, Spectrometers, Detectors and Associated Equipment},
  835:186--225, 2016.

\bibitem{Haffke_2011}
M.~Haffke, L.~Baudis, T.~Bruch, A.D. Ferella, T.~Marrodán Undagoitia,
  M.~Schumann, Y.-F. Te, and A.~van~der Schaaf.
\newblock Background measurements in the gran sasso underground laboratory.
\newblock {\em Nuclear Instruments and Methods in Physics Research Section A:
  Accelerators, Spectrometers, Detectors and Associated Equipment},
  643(1):36–41, July 2011.

\bibitem{JRC135812}
Paepen J, Van~Ammel R, Hult M, Marissens G, Stroh H, Emteborg H, and Seghers J.
\newblock Normconstruct: Results of the collaborative assessment experiment for
  the validation of the precision of cen/ts 17216.
\newblock (KJ-02-24-271-EN-N (online)), 2024.

\bibitem{reback2020pandas}
The pandas~development team.
\newblock pandas-dev/pandas: Pandas (v2.2.3).
\newblock https://doi.org/10.5281/zenodo.13819579, September 2024.

\bibitem{Hunter:2007}
J.D. Hunter.
\newblock Matplotlib: A 2d graphics environment.
\newblock {\em Computing in Science \& Engineering}, 9(3):90--95, 2007.

\bibitem{harris2020array}
Charles~R. Harris et~al.
\newblock Array programming with {NumPy}.
\newblock {\em Nature}, 585(7825):357--362, Sep 2020.

\bibitem{2020SciPy-NMeth}
{SciPy 1.0 Contributors}.
\newblock {{SciPy} 1.0: Fundamental Algorithms for Scientific Computing in
  Python}.
\newblock {\em Nature Methods}, 17:261--272, 2020.

\bibitem{Kluyver2016jupyter}
Thomas Kluyver et~al.
\newblock Jupyter notebooks -- a publishing format for reproducible
  computational workflows.
\newblock In F.~Loizides and B.~Schmidt, editors, {\em Positioning and Power in
  Academic Publishing: Players, Agents and Agendas}, pages 87 -- 90. IOS Press,
  2016.

\end{thebibliography}

\appendix


\end{document}